\documentclass[aps,pre,twocolumn,nofootinbib,tightenlines,superscriptaddress,nopacs,amsmath,amssymb,final,letterpaper]{revtex4-1}

\usepackage[utf8]{inputenc}
\usepackage{calc}
\usepackage{graphicx}
\usepackage{amsmath,amssymb,amsthm}
\usepackage{bbold}
\usepackage{bm}
\usepackage{color}
\usepackage{hyperref}
\usepackage[dvipsnames]{xcolor}

\hypersetup{bookmarks=true, unicode=false, pdftoolbar=true, pdfmenubar=true, pdffitwindow=false, pdfstartview={FitH}, pdfnewwindow=true, colorlinks=true, linkcolor=blue, citecolor=blue, filecolor=magenta, urlcolor=blue}

\begin{document}

\title{Hypergraph reconstruction from network data}

\author{Jean-Gabriel \surname{Young}}
\email[Corresponding author: ]{jean-gabriel.young@uvm.edu}
\affiliation{Center for the Study of Complex Systems, University of Michigan, Ann Arbor, Michigan 48109, USA}
\affiliation{Department of Mathematics and Statistics, University of Vermont, Burlington, VT 05404, USA}
\affiliation{Vermont Complex Systems Center, University of Vermont, Burlington, VT 05404, USA} 
\author{Giovanni \surname{Petri}}
\affiliation{ISI Foundation, Via Chisola 5, Torino 10126, Italy}
\author{Tiago P. \surname{Peixoto}}
\affiliation{Department of Network and Data Science, Central European University, 1100 Vienna, Austria}
\affiliation{ISI Foundation, Via Chisola 5, Torino 10126, Italy}
\affiliation{Department of Mathematical Sciences, University of Bath, Bath BA2 7AY, United Kingdom}

\date{\today}
\begin{abstract}
Networks can describe the structure of a wide variety of complex systems by specifying which pairs of entities in the system are connected.
While such pairwise representations are flexible, they are not necessarily appropriate when the fundamental interactions involve more than two entities at the same time. 
Pairwise representations nonetheless remain ubiquitous, because higher-order interactions are often not recorded explicitly in network data.
Here, we introduce a Bayesian approach to reconstruct latent higher-order interactions from ordinary pairwise network data. 
Our method is based on the principle of parsimony and only includes higher-order structures when there is sufficient statistical evidence for them.
We demonstrate its applicability to a wide range of datasets, both synthetic and empirical.
\end{abstract}

\maketitle

\section{Introduction}

Empirical networks are often locally dense and globally sparse \cite{williamson2018random}.
Whether they are social, biological, or technological~\cite{newman2018networks}, they comprise large groups of densely interconnected nodes, even when only a small fraction of all possible connections exist.
This situation leads to delicate modeling challenges: How can we account for two seemingly contradictory properties of networks---density and sparsity---in our models?

Abundant prior work going back to the early days of social network analysis~\cite{frank1986markov,iacobucci1990social} and network science~\cite{watts2002identity,newman2003properties} suggests that higher-order interactions~\cite{battiston2020networks} are a possible explanation for the local density of networks \cite{newman2003properties,williamson2018random}.
According to this reasoning, entities are connected because they have a shared context---a higher-order interaction---within which connections can be created~\cite{latapy2008basic}.
It is clear that a phenomenon along these lines occurs in many social processes: scientists appear as collaborators in the Web of Science because they co-author papers together; colleagues exchange emails because they are part of the same department or the same division of a company.
It is also known that similar phenomena explain tie formation in a broader range of networked systems, including biological, technological or informational systems \cite{battiston2020networks}.

The ubiquity of higher-order interactions provides a simple and universal explanation for the observed structure of empirical networks.
If we assume that most ties are created within contexts of limited scopes, then the resulting networks are locally dense, matching empirical observations ~\cite{pollner2005preferential,hebert2015complex,williamson2018random}.

Despite their tremendous explanatory power, higher-order interactions are seldom used directly to model empirical systems, due to a lack of data~\cite{battiston2020networks}.
Indeed, while the context is directly observable for some systems---say, co-authored papers or co-locating species---it is unavailable for several others, including brain data~\cite{white1986structure}, typical social interaction data~\cite{atkin1974book}, and ecological competitor data~\cite{grilli2017higher} to name only a few.

\begin{figure}
\centering
\includegraphics[width=0.8\linewidth]{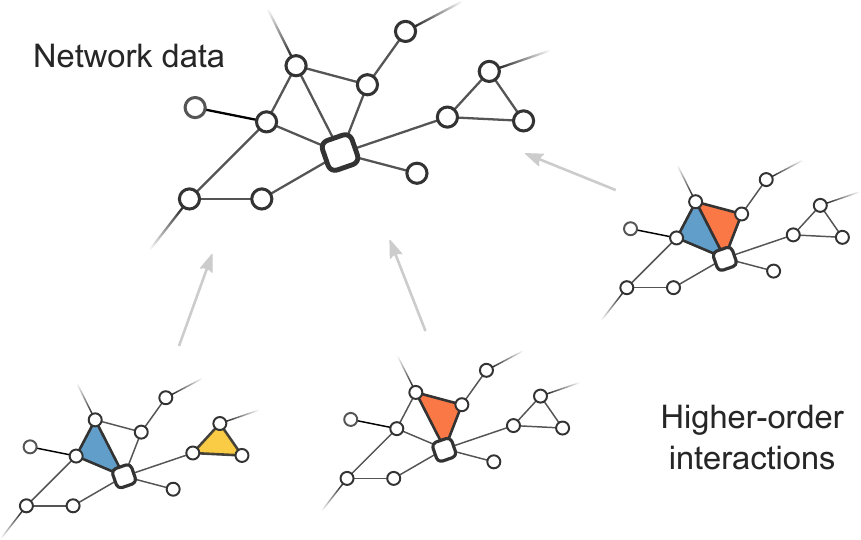}
\caption{
    \textbf{Projected higher-order interactions.}
    We show the extended ego network (circle nodes) of a participant (square node) of the \mbox{AddHealth} study~\cite{resnick1997protecting}.
    Friendships are measured between pairs of participants (links and nodes respectively), even when the fundamental units are groups of friends \cite{atkin1974book}.
    Multiple combinations of groups and isolated friendships lead to the same network (gray arrows).
}
\label{fig:addhealth_example}
\end{figure}

As a specific motivating example, consider one of the empirical social networks gathered as part of the US National Longitudinal Study of Adolescent to Adult Health~\cite{resnick1997protecting}.
This dataset is constructed using surveys, where participants are asked to nominate their friends.
Even though there are good reasons to believe that people often interact because of higher-order groups~\cite{atkin1974book}, the survey cannot reveal these groups as it only inquires about pairwise relationships.
If we actually need the higher-order interactions to give an appropriate description of the social dynamics at play~\cite{atkin1974book}, what should we do with such inadequate survey data?
As we show in Fig.~\ref{fig:addhealth_example}, there are many kinds of higher-order interactions that are compatible with the same network data.
How can one pick among all these possible higher-order descriptions?

Prior work on higher-order interaction discovery in network data often uses cliques---fully connected subgraphs---to identify the interactions~\cite{patania2017topological,petri2013networks,petri2013topological}.
Clique-based methods are straightforward to implement because they rely on clique enumeration, a classical problem for which we have exact~\cite{bron1973algorithm,tomita2006worst} and sampling~\cite{jain2017fast} algorithms that work well in practice.
However, clique decompositions do not offer a satisfactory solution to the recovery problem alone.
Networks typically admit many possible clique decompositions, which begs the question of which one to pick.
For example, a triangle can be decomposed as a single 2-clique, or as three 1-cliques (i.e., as edges); see Fig.~\ref{fig:addhealth_example}.
In general, the multiplicity of possible solutions implies that higher-order interaction recovery is an ill-posed inverse problem.
It becomes well-posed only once we add further constraints on what constitutes a good solution.
Thus, existing approaches have sought to address  the ill-posed nature of the higher-order interaction recovery problem in various indirect ways. 
For instance, in graph theory, it is customary to look for a minimal set of cliques covering the network~\cite{erdos1966representation,coutinho2020covering}. 
Other methods appeal to notions of randomness and generative modeling to regularize the problem~\cite{wegner2014subgraph,williamson2018random,koutra2014vog,liu2018reducing}. 
These methods describe an explicit process by which one goes from higher-order data to networks, and can therefore assign a likelihood to possible higher-order data representations, allowing the user to single out representations.

In the present work, we develop a Bayesian method for the inference of higher-order of interactions from network.
Given a network as input, the method identifies the parts of the network best explained by latent higher-order interactions.
Our approach is based on the principle of parsimony and directly addresses the ill-posedness of the reconstruction problem with the methods of information theory.
We show that the method can finds compact descriptions of many empirical networked systems by using latent higher-order interactions, thereby demonstrating that such interactions are in complex systems.

\section{Results and discussion}

\subsection{Generative model}
\label{subsec:model}

The problem we solve is illustrated in Fig.~\ref{fig:addhealth_example}.
We have a system we believe is best described with higher-order interactions, but we can only view its structure through the lens of pairwise measurements (an undirected and simple network $G$); our goal is to reconstruct these higher-order interactions from $G$ only.

For convenience, we encode the higher-order interactions with a hypergraph $H$~\cite{torres2020and}.
We represent a higher-order interaction between a set of $k$ nodes $i_1,..,i_k$ with a hyperedge of size $k$.
Empirical data often contain repeated interactions between the same group of nodes, so we use hypergraphs with repeated hyperedges and encode the number of hyperedges connecting nodes $i_1,..,i_k$ as $A_{i_1,..,i_k}\geq 0$.

Our method then makes use of a Bayesian generative model to deduce one such hypergraph $H$ from some network dataset $G$.
This generative model gives an explicit description of how the network data $G$ is generated when there are latent higher-order interactions $H$. With a generative model in place, we can compute the posterior probability
\begin{equation}
    \label{eq:bayes} P(H|G) =  \frac{P(G|H)P(H)}{P(G)}
\end{equation}
that the latent hypergraph is $H$, given the observed network $G$. 
In this equation, $P(G|H)$ and $P(H)$ define our generative model for the data, and its evidence $P(G)=\sum_H P(G|H)P(H)$ functions as a normalization constant.

The appeal of such a Bayesian generative formulation is that we can use $P(H|G)$ to make queries about the hypergraph $H$. 
What was the most likely set of higher-order interactions? What is the probability that a particular interaction was present in $H$ based on $G$? How large were the latent higher-order interactions?
All of the queries can be answered by computing appropriate averages over $P(H|G)$.
As is made evident by Eq.~\eqref{eq:bayes}, however, we first have to introduce two probability distributions so that we may compute $P(H|G)$ at all.
We now define these distributions in detail.

\begin{figure*}
    \centering
    \includegraphics[width=0.8\linewidth]{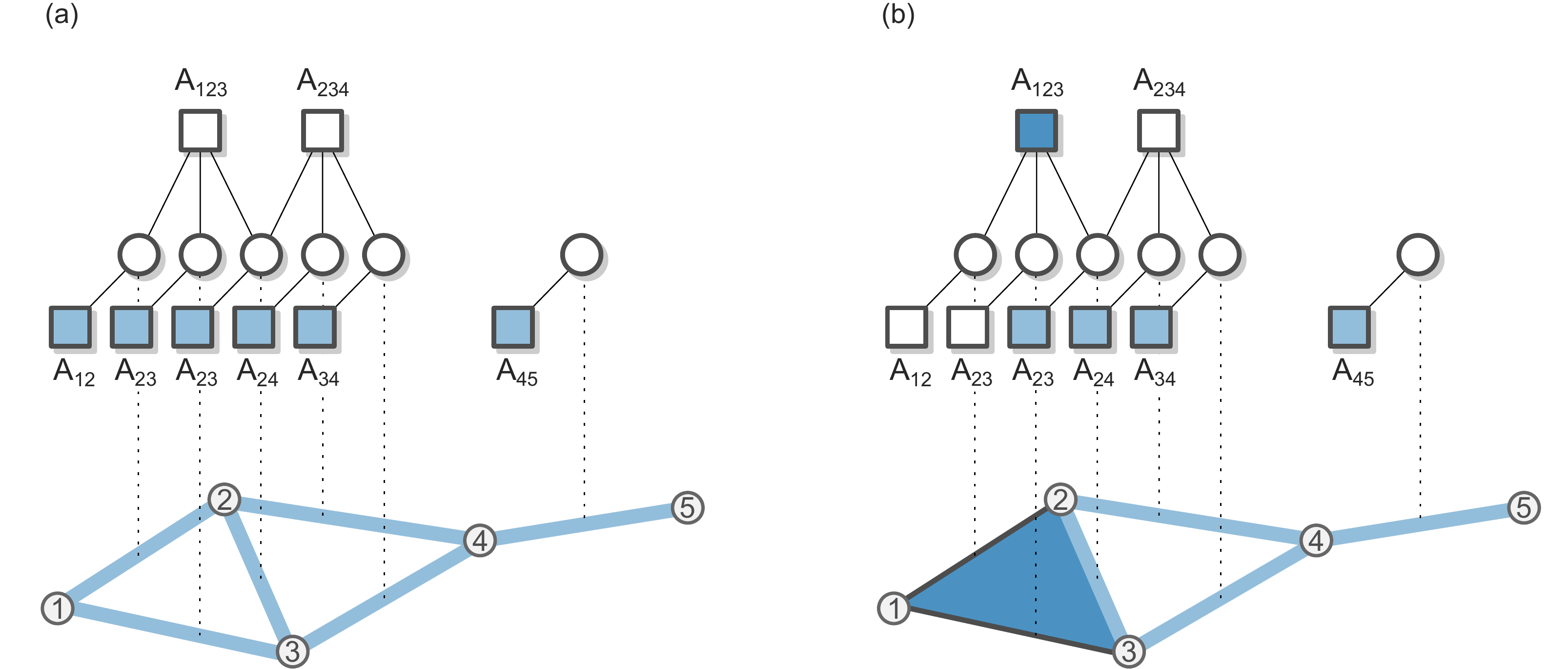}
    \caption{
      \textbf{Encoding hypergraphs as factor graphs.}
      The two panels show the factor graph encoding of two different hypergraphs that project to the same network, shown at the bottom of the figure.
      \textbf{(a)} A hypergraph without higher-order interactions is obtained by associating each edge in $G$ (blue lines) to a node in the factor graph (empty circles); this correspondence is illustrated with a dashed line. Each node in the factor graph is then connected to a factor $A_{ij}$ (squares) where $(i,j)$ is the corresponding edge in $G$. Higher-order factors $A_{ijk}$ corresponding to possible hyperedges of 3 nodes are also added, and connected to the edges $(i,j), (j,k), (i,k)$ they comprise. Since there is no $4$-clique in the graph, the construction stop at this step. Particular combination of hyperedges can then be specified by activating some factors (coloring them blue), here all the factors corresponding to the edges.  \textbf{(b)} To encode a hypergraph $H$ with one higher-order interaction, we mark the factor $A_{123}$ as active.
    }
    \label{fig:factor_graph}
\end{figure*}

\subsubsection{Projection component}
\label{subsubsec:projection}

The first distribution, $P(G|H)$, is called the projection component of the model.
It tells us how likely a particular network $G$ is when the latent hypergraph $H$ is known.

We use a direct projection component and deem two nodes connected in $G$ if and only if these nodes jointly appear in any of the hyperedges of $H$.

This modeling choice is broadly applicable.
For instance, when researchers measure the functional connectivity of two brain regions, they record a connection irrespective of whether the regions peaked as a pair or as jointly with many other regions.
Likewise, surveyed social networks  contain records of friendships that can be attributed to interactions between pairs of individuals, and to interactions that arise from larger groups.

Certain authors use more nuanced projection components~\cite{barber2012clique,williamson2018random} and do not assume that the joint participation of two nodes in a hyperedge necessarily leads to measured pairwise interactions. 
As we have argued in the introduction, we believe that such components blur the lines between community detection and higher-order interaction reconstruction.
Hence, we treat measurement as a separate issue to be handled with the methods of Refs.~\cite{young2020bayesian,peixoto2018reconstructing}, for instance.

We formalize the projection component as follows.
We set $P(G|H)=1$ only when (i) each pair of nodes connected by an edge in $G$ appears jointly in at least one hyperedge of $H$, and (ii) no two disconnected nodes of $G$ appear together in a hyperedge of $H$.
If either of these conditions is violated, then we set $P(G|H)=0$.
We can express this definition mathematically as
\begin{equation}
    \label{eq:likelihood}
    P(G|H) = \left\{
    \begin{array}{ll}
       1 & \text{if }G=\mathcal{G}(H),\\
       0 & \text{otherwise.}
    \end{array}
    \right.
\end{equation}
where we use $\mathcal{G}(H)$ to denote the projection of $H$  and use $G=\mathcal{G}(H)$ to say that $H$ projects to $G$, or equivalently that (i) and (ii) hold.

Testing $G\stackrel{?}{=}\mathcal{G}(H)$ might appear unwieldy at first but, thankfully, a factor graph encoding of $H$ can help us compute the projection component efficiently by highlighting existing relationships between the edges and cliques of $G$~\cite{bishop2006pattern}.

To construct this factor graph, we begin by creating two separate sets of nodes: one representing the edges of $G$, and one representing the cliques of $G$.
Crucially, the second set contains a node for every clique of $G$, even the included ones like the edges of a triangle, the triangles of a 4-clique, and so on.
We call this set the set of factors and refer to nodes in the first set simply as nodes.
We obtain a factor graph, by connecting a factor and a node when the corresponding clique contains the corresponding edge.

This construction is illustrated in Fig.~\ref{fig:factor_graph} for a simple graph of 5 nodes.
In Fig.~\ref{fig:factor_graph}, we see that, for example, the edge between nodes $1$ and $2$ is part of the triangle $\{1,2,3\}$ in $G$, and it is therefore connected to the factor $A_{123}$.
This edge is also part of the 2-clique $\{1,2\}$ so it is connected to the factor $A_{12}$, too.

The resulting factor graphs can encode particular hypergraphs $H$ by assigning integers to the factors, corresponding to the number of times every hyperedge appears in $H$.
For example, by setting $A_{123}=1$ and $A_{23}=A_{24}=A_{34}=A_{45}=1$, we can encode a hypergraph with five hyperedges, one of size 3 and four of size 2; see Fig.~\ref{fig:factor_graph}b.
We obtain a simple graph representation of the same data by setting $A_{123}=0$ and $A_{12}=1$ instead; see Fig.~\ref{fig:factor_graph}a.

It is straightforward to check whether $G=\mathcal{G}(H)$ holds with this encoding.
The first condition---all the connected nodes of $G$ are connected by at least one hyperedge in $H$---can be verified by checking that every node of the factor graph is connected to at least one active factor, defined as $A_{i_1,..,i_k}>0$.
The second condition---no pairs of disconnected nodes in $G$ are connected by a hyperedge of $H$---is always satisfied by construction, because no factor connects two disconnected nodes of $G$, so we never represent these forbidden hyperedges with our factor graph.

We note that the factor graph can be stored relatively efficiently, by first enumerating the maximal cliques---cliques not included in larger cliques---and then constructing an associative array indexed by cliques, which we expand only when included cliques are needed.
Even though enumerating maximal clique is technically an \textsc{NP}-hard problem~\cite{karp1972reducibility}, state-of-the-art enumeration algorithms tend to work well on sparse empirical network data~\cite{bron1973algorithm,tomita2006worst,fox2020finding}, and indeed we have found that enumeration is not problematic in our experiments.

\subsubsection{Hypergraph prior}

The second part of Eq.~\eqref{eq:bayes}, $P(H)$, is the hypergraph prior.
Empirical hypergraphs generally have a few properties that a reasonable prior should account for~\cite{aksoy2020hypernetwork}: the size of interactions vary; some of these interactions are repeated, and not all nodes are connected by a hyperedge.
It turns out that an existing model~\cite{darling2005structure}, known as the Poisson Random Hypergraphs Model (PRHM), reproduces all of these properties.
Hence, we adopt it as our hypergraph prior.
The PRHM was initially developed to study critical phenomena in hypergraphs~\cite{darling2005structure}; here, we use it to make posterior inferences about networks.

In a nutshell, the PRHM stipulates that the number of hyperedges connecting a set of nodes is a random variable, whose mean $\lambda_k$
only depends on the size $k$ of the set.
The variable follows a Poison distribution, such that the number of hyperedges connecting the nodes $i_1,..,i_k$ equals to $A_{i_1,..,i_k}$ with probability
\begin{align}
    \label{eq:indiv_edge_prob}
    P(A_{i_1,..,i_k}|\lambda_k) &= \frac{\lambda_k^{A_{i_1,..,i_k}}}{A_{i_1,..,i_k}!}e^{-\lambda_k},
\end{align}
where $A_{i_1,..,i_k}$ is invariant with respect to permutation of the indexes.
The PRHM also models all the hyperedges as independent.
Hence, the probability of a particular hypergraph can be calculated as
\begin{align}
    P(H|\bm{\lambda}) &= \prod_{k=2}^{L} \ \prod_{i_1,..,i_k\in C_k^N} P(A_{i_1,..,i_k}|\lambda_k)\notag\\
                      &= \prod_{k=2}^{L} \ \prod_{i_1,..,i_k\in C_k^N} \frac{\lambda_k^{A_{i_1,..,i_k}}}{A_{i_1,..,i_k}!}e^{-\lambda_k},
    \label{eq:hypergraph_prob}
\end{align}
where $L$ is the maximal hyperedge size, $C_k^N$ denotes all possible subsets of size $k$ of $\{1,...,N\}$, and where $\bm{\lambda}$ refers to all the rates collectively.

Equation~\eqref{eq:hypergraph_prob} expresses the probability of $H$ in terms of individual hyperedges.
To obtain a simpler form, we notice that the number $E_k$ of hyperedges of size $k$, can be calculated as
\begin{equation}
    E_k=  \sum_{i_1,..,i_k\in C_k^N}  A_{i_1,..,i_k}
\end{equation}
and that there are precisely $\binom{N}{k}$ terms in the product over all sets of nodes of size $k$.
We can use these simple observations to rewrite Eq.\eqref{eq:hypergraph_prob} as
\begin{align}
    P(H|\bm{\lambda}) = \prod_{k=2}^{L} \frac{\lambda_k^{E_k} e^{-\binom{N}{k}\lambda_k}}{Z_k},
    \label{eq:hypergraph_prob_simplified}
\end{align}
where we have defined 
\begin{equation}
    \label{eq:Z_k}
    Z_k= \prod_{i_1,..,i_k\in C_k^N} A_{i_1,..,i_k}! = \prod_{m=1}^{\infty} (m!)^{\eta_m^{(k)}},
\end{equation}
and where $\eta_m^{(k)}$ is the number of hyperedges of size $k$ that are repeated precisely $m$ times.

In this form, it is clear that the parameters $\bm{\lambda}$ control the density of $H$ at all scales.
Hence, they more or less determine the kind of hypergraphs we expect to see a priori, and therefore have a major effect on the model output.
How can we choose these important parameters carefully?

We propose to a hierarchical empirical Bayes approach, in which we treat $\bm{\lambda}$ as unknowns themselves drawn from prior distributions.
We use a maximum entropy, or least informative, prior for $\bm{\lambda}$, because we have no information whatsoever about $\bm{\lambda}$ a priori. 
The only thing we know is that these parameters take values in $[0,\infty)$ and are modeled with a finite mean~\cite{darling2005structure}.
Hence, the maximal entropy prior of interest is the exponential distribution
\begin{equation}
    \label{eq:max_ent}
   P(\lambda_k|\nu_k) =  \frac{e^{-\lambda_k/\nu_k}}{\nu_k},
\end{equation}
of mean $\nu_k$.
We obtain a complete hyperprior for $\bm{\lambda}$ by using independent priors for all sizes $k$, $P(\bm{\lambda}|\bm{\nu}) = \prod_{k=2}^L P(\lambda_k|\nu_k)$. 
Integrating over the support of $\bm{\lambda}$, we find that the prior for $H$ is now
\begin{align}
    P(H| \bm{\nu})  &= \int P(H|\bm{\lambda}) P(\bm{\lambda}|\bm{\nu}) d\bm{\lambda},\notag\\
                    &= \prod_{k=2}^{L} \frac{E_k!}{Z_k \nu_k} \left[\frac{1}{\nu_k} +\binom{N}{k}\right]^{-(E_k+1)},
    \label{eq:hypergraph_prior_generic}
\end{align}
with $\bm{\nu}$ fixed.

It might appear that we have only pushed our problem further ahead---we got rid of $\bm{\lambda}$ but we now have a whole new set of parameters on our hands. 
Notice, however, that the new parameters $\bm{\nu}$ do not have as direct an effect on $H$.
A whole range of densities is now compatible with any choice of $\bm{\nu}$.
And as a result, the model can assign significant probabilities to hypergraphs that project to networks of the correct density, even when the hyperprior is somewhat in error.
Hence, we safely fix the new parameters $\bm{\nu}$ with empirical Bayes without risking strongly biased results.

With these precautions in place, we use the observed number of edges $E$ in the network $G$ to choose $\bm{\nu}$.
Our strategy is to equate $E$ to the expected number of edges $\langle E(\bm{\nu}) \rangle$ in the network $\mathcal{G}(H)$ obtained by projecting $H$ drawn from $P(H|\bm{\nu})$. 
This expected density can be approximated as
 \begin{equation}
  \label{eq:approx_density}
  \langle E(\bm{\nu}) \rangle = \sum_{k=2}^{L}\nu_k \binom{N}{k}
\end{equation}
by assuming that hyperedges do not overlap on average.
To set the individual values of $\nu_k$, we further require that all sizes contribute equally to the final density, with $\nu_k \binom{N}{k} = \mu$ for some constant $\mu$.
Substituting these equalities in Eq.~\eqref{eq:approx_density}, we obtain
\begin{equation}
    \label{eq:mu_approx_prior}
    \mu =  E/(L-1),
\end{equation}
such that the prior in Eq.~\eqref{eq:hypergraph_prior_generic} becomes
\begin{equation}
    \label{eq:approx_prior}
    P(H) = \prod_{k=2}^{L} \frac{E_k!}{Z_k \mu \binom{N}{k}^{E_k}} \left[\frac{1}{\mu} +1\right]^{-(E_k+1)},
\end{equation}
which is the equation we will use henceforth, with $\mu=E/(L-1).$

\subsection{Properties of the posterior distribution}
\label{subsec:properties}

The model defined in Eqs.~\eqref{eq:likelihood} through \eqref{eq:approx_prior} has two crucial properties.

The first noteworthy property is that the model assigns a higher posterior probability to hypergraphs without repeated hyperedges, even though the prior $P(H)$ allows for duplicates.
An explicit calculation of how $P(H|G)$ scales with the number of duplicated hyperedges can illustrate this fact.
Indeed, consider a hypergraph $H_0$ with no repeated hyperedges, for which $P(G|H_0)=1$.
Write as $\alpha$ the fraction of $k$-cliques connected by a hyperedge in $H_0$, and consider an experiment in which an average of $\beta\geq0$ additional hyperedges are placed on top of the hyperedges of size $k$ already present in $H_0$.
In these hypergraphs, the expected number of hyperedges of size $k$ is $E_k=\alpha(1+\beta)\binom{N}{k}$ and $\log Z_k$ is approximated by
\begin{equation*}
  \sum_{i_1,...,i}  \log A_{i_1,...,i}! \approx \alpha \binom{N}{k} \log(1+\beta)!,
\end{equation*}
see Eq.~\eqref{eq:Z_k}.
Substituting our various formula in the logarithm of $P(H)$, and using the Stirling approximation $\log n ! \approx n \log n - n$, we find that 
\begin{equation*}
    \log P(H)\sim - \alpha(1+\beta)\binom{N}{k}\log\left[\frac{1 + \frac{1}{\mu}}{\alpha}\right]
\end{equation*}
This equation tells us that the log-posterior $\log P(H|G)$ decreases with growing $\beta$, because the argument of the logarithm is greater or equal to one.
Furthermore the likelihood equals one by construction, which implies that the scaling of the prior determines the scaling of the posterior.
Hence, the hypergraphs $H$ generated by adding duplicated hyperedges to $H_0$---that is by increasing $\beta$---are less likely than $H_0$.

A second noteworthy property of the model is that it favors sparser hypergraphs: as long as $P(G|H)=1$, the fewer hyperedges, the better.
To make this observation precise, suppose we have a hypergraph $H_m$ that can be termed minimal for $G$: every edge of $G$ is covered by exactly one hyperedge of $H_m$ and no more.
We observe that we cannot improve on the posterior probability of $H_m$ by adding a hyperedge, even when this new hyperedge does not fully repeat an existing one.
Indeed, consider the hypergraph $H'_m$ created by adding a hyperedge of size $k$ to $H_m$.
(For example, we could add a hyperedge of size $3$ on a triangle whose sides were already covered by edges, but did not yet participate in any larger hyperedge together.)
By direct calculation, the ratio of posterior probability for $H'_m$ and $H_m$ equals
\begin{equation*}
  \frac{P(H'_m|G)}{P(H_m|G)} = \frac{Z_k}{Z_k'}\frac{E_k + 1}{\binom{N}{k}}\left[\frac{1}{\mu}+1\right]^{-1},
\end{equation*}
where $Z_k'$ is the quantity in Eq.~\eqref{eq:Z_k} for the modified minimal hypergraph, and $Z_k$ is the same quantity for the minimal hypergraph.
One can show that this ratio is always smaller than one and that, as a result, adding a spurious hyperedge to a minimal hypergraph decreases the posterior probability.
The proof is straightforward and relies on the observation that for a minimal hypergraph, we have $E_k\leq \binom{N}{k}$, $Z_k=1$, and $Z_k'=1$ or $Z_k'=2$.
The result follows by direct computation when $E_k<\binom{N}{k}$ and uses the fact that that $Z_k'=2$ when $E_k=\binom{N}{k}$ (because adding a single hyperedge to a completely connected minimal hypergraph means one has to double-up one hyperedge).

As a corollary of the two above observations, we conclude that the minimal hypergraphs are high-quality local maxima of $P(H|G)$.
We cannot simply pick one of these optima as our reconstruction, however, because there may exist multiple ones of comparable quality.
Further, non-optimal hypergraphs may account for a significant fraction of the posterior probability in principle.
Instead, we handle these possibly conflicting descriptions by combining them.

\subsection{Posterior estimation}
\label{subsec:posterior_estimation}

In the Bayesian formulation of hypergraph inference, estimating a given quantity of interests always amount to computing an expectations over the posterior distribution $P(H|G)$.
For example, the expected number of hyperedges of size $k$ can be computed as  $\langle E_k\rangle=\sum_{H} E_k(H) P(H|G).$
More generally, we are interested in averages of the form
\begin{equation}
    \label{eq:posterior_average} 
    \langle f(H) \rangle = \sum_{H} f(H) P(H|G)
\end{equation}
for arbitrary functions $f$ that map hypergraphs to vectors or scalars.

The summation in Eq.~\eqref{eq:posterior_average} is unfortunately intractable: the set of possible hypergraphs grows exponentially in size with both the number of nodes and the maximal size of the hyperedges.
Hence, we propose a Markov Chain Monte-Carlo (MCMC) algorithm to evaluate Eq.~\eqref{eq:posterior_average}.
This kind of approach generates a random walk over the space of all hypergraphs, with a limiting distribution identical to $P(H|G)$.
We use the Metropolis-Hastings (MH) construction to implement the random walk.
As is usual, the algorithm consists of proposing a move from $H$ to $H'$ with probability $Q(H\ \leftarrow H')$ and accepting it with probability~\cite{andrieu2003introduction}
\begin{align}
    a &= \min\left\{1,  \frac{Q(H\ \leftarrow H')}{Q(H'\leftarrow H\ )}\frac{P(H'|G)}{P(H\ |G)}\right\},\notag\\
      &= \min\left\{1,  \frac{Q(H\ \leftarrow H')}{Q(H'\leftarrow H\ )}\frac{P(G|H')P(H')}{P(G|H\ )P(H\ )}\right\}.
    \label{eq:acceptance_prob}
\end{align}

We use the factor graph representation of $H$ to define these Monte Carlo moves this encoding facilitates checking the value of $P(G|H')$.
Hence, we can state the moves as modifications to the value of the factors $A_{i_1,....i_{k}}$, i.e., the number of hyperedges connecting particular sets of nodes.

The specific set of moves we use goes as follow.
For every move, we begin by choosing a maximal factor node uniformly at random from the set of all such factors.
We select a size $\ell$ uniformly at random from $\{2,3,...,k\}$ where $k$ is the size of the clique corresponding to the current maximal  factor.
Then we select one of the subfactors $A_{i_1,..,i_{\ell}}$ of size $\ell$ uniformly at random, among the $\binom{k}{\ell}$ factors of that size, and we update the selected factors as either $A_{i_1,..,i_{\ell}}'=A_{i_1,..,i_{\ell}} + 1$ or $A_{i_1,..,i_{\ell}} '= A_{i_1,..,i_{\ell}} - 1$ (with probability $1/2$).
If $A_{i_1,..,i_{\ell}}$ was already equal to zero, we force $A_{i_1,..,i_{\ell}} '= A_{i_1,..,i_{\ell}} + 1$.
Therefore we have that \begin{equation}
  \frac{Q(H\ \leftarrow H')}{Q(H'\leftarrow H\ )} =
  \begin{cases}
    1 & \text{ if } A_{i_1,..,i_{\ell}} > 0,\\
    1/2 & \text{ if } A_{i_1,..,i_{\ell}} = 0,\\
    2 & \text{ if } A_{i_1,..,i_{\ell}}' = 0.\\
  \end{cases}
\end{equation}
Finally, we check whether $P(G|H)=1$ using the factor representation, and compute the ratio $P(H')/P(H)$ to obtain the acceptance probability $a$.
We test for acceptance and, if the move is accepted, we record the update.
Otherwise, we do nothing.

The posterior distribution is rugged so the initialization of the MCMC algorithm matters a great deal in practice.
Building on our observations about the properties of $P(H|G)$, we select as our initialization the hypergraph with one hyperedge for every maximal clique of $G$.
This starting point is not a known optimum of $P(H|G)$ but it is close to many of them.
Hence, chains initialized at this point have a fairly good chance of converging to a good optimum.
And indeed, in our experiments, we find that the maximal clique initialization works much better than a random initialization, an edge initialization or an empty one.

\subsection{Recovery of planted higher-order interactions in synthetic data}
\label{subsec:recovery}

\begin{figure*}   
    \includegraphics[width=0.66\linewidth]{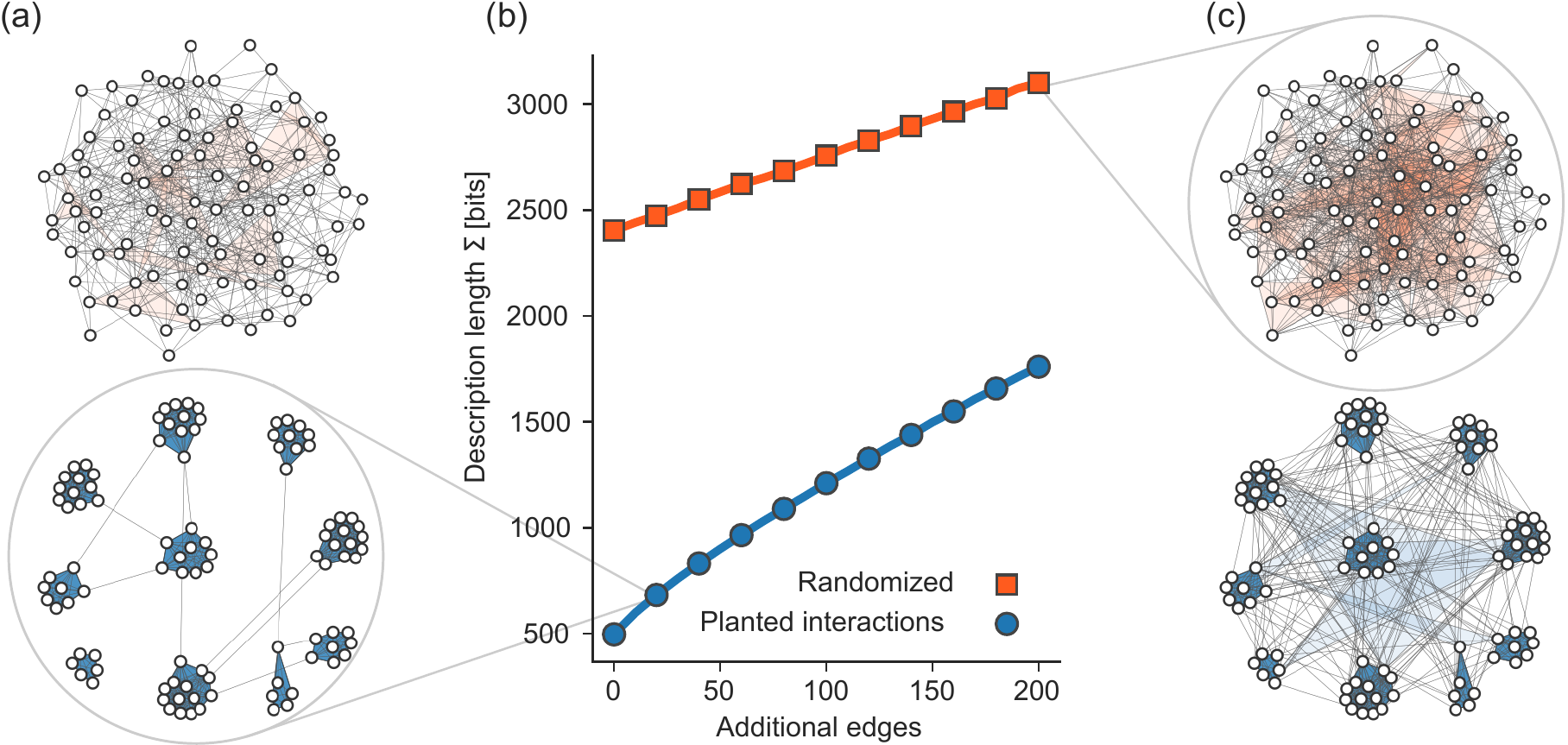} 
    \caption{
        \textbf{Reconstruction in random networks with and without higher-order interactions.} 
        \textbf{(a)} Two networks, one obtained by projecting a hypergraph of $10$ disjoint hyperedges of unequal sizes (blue shades), and the other obtained by drawing uniformly from all networks with the same number of edges (orange shades).
        \textbf{(b)} Description length $\Sigma$  of the networks as a function of the number of additional edges, chosen uniformly at random from the set of non-edges.
        \textbf{(c)} The two networks of panel (a), with 200 additional edges.
        In panel (a) and (c), we show the group interactions uncovered by our method with shaded colors.
        The description lengths are averaged over 10 independent realizations of the network generation and inference processes.
        Error bars of 1 standard deviations are too narrow to see with the naked eye.
    }
    \label{fig:planted_clique}
\end{figure*}

To develop an intuition for the workings of our method, we first use our algorithm to uncover higher-order interactions in synthetic data generated by the model appearing in Eq.~\eqref{eq:likelihood}--\eqref{eq:approx_prior}, altered slightly to facilitate the interpretation of the results.
In this experiment, we create a hypergraph that comprises a few large disconnected hyperedges, and we add several random edges (chosen uniformly from the set of all edges) to create a noisy hypergraph  $\tilde{H}$.
We then project this noise hypergraph to obtain a network $\mathcal{G}(\tilde{H})$, which we feed to our recovery algorithm as input.
Our goal in this experiment is to find the hypergraph $H^*$ that maximizes the posterior probability $P(H|\mathcal{G}(\tilde{H}))$ (we do not use the full samples given by our MCMC algorithm just yet).
We can consider the experiment successful if $H^*$ contains all the higher-order interactions planted in $\tilde{H}$.

The results of this experiment are reported in Fig.~\ref{fig:planted_clique}.
At the bottom of Fig.~\ref{fig:planted_clique}a, we show a typical example of what the projected networks $\mathcal{G}(\tilde{H})$ look like when there are very few added random edges.
In this regime, the recovered higher-order interactions (in blue) correspond perfectly to those planted in $\tilde{H}$,
for the entire range of parameters we investigated.
For the sake of comparison, we also generate an equivalent random network, obtained by completely rewiring the edges of $\mathcal{G}(\tilde{H})$, see the top of Fig.~\ref{fig:planted_clique}a.
(Equivalently, we generate an Erd\H{o}s-R\'enyi graph with an equal number of edges~\cite{erdHos1960evolution}.)
This network has the same number of edges as $\mathcal{G}(\tilde{H})$ but is otherwise unstructured. 
As expected, we find no higher-order interactions beyond the random triangles that occur at this density~\cite{bollobas1976cliques}.

If we add many more random edges, we obtain the results shown in Fig.~\ref{fig:planted_clique}c.
Again, we can recover the planted higher-order interactions, but we also start to find additional ones, due to the appearance of random triangles formed by triplets of edges added at random~\cite{shi2007networks}.
To understand this behavior we turn to the  minimum description length (MDL) interpretation of Bayesian inference~\cite{mackay2003information,grunwald2007minimum}.

In a nutshell, the description length is the number of bits that a receiver and a sender with shared knowledge of the model $P(G,H)$ would need to communicate the network $G$ to one another.
This communication costs can be minimized by finding a hypergraph $H^*$ that is both cheap to communicate and projects to $G$; receivers who know $P(G,H)$ also know that they  can project  $H^*$ to find $G=\mathcal{G}(H)^*$.
From this communication perspective, hypergraphs with as few hyperedges as possible are good candidates because they are cheaper to send \cite{koutra2014vog}.
The connection with Bayesian inference is that $H^*$ happens to coincide with the hypergraph which maximizes the posterior probability $P(H|G)$ (see Supplementary Notes 1 and 2 for a detailed discussion).
Hence, maximum a posteriori inference is equivalent to compression.

Reviewing our experiment with the MDL interpretation in mind illuminates the results.
In Fig.~\ref{fig:planted_clique}b, we plot the description length provided by our model, for levels of randomness that interpolate between the easy regime shown in Fig.~\ref{fig:planted_clique}a, and the much more random regime appearing in Fig.~\ref{fig:planted_clique}c.
We find that the model compresses those networks that have planted interactions much better than their randomized equivalents. 
These results make intuitive sense: networks with planted interactions contain large cliques, and these cliques can be harnessed to communicate regularities in $G$.
As can be expected, these savings disappear once the large cliques are destroyed by rewiring.

\subsection{Recovery of planted higher-order interactions in empirical data}
\label{subsec:recovery}

Having verified that the method works when the higher-order network possess little structure beyond disjoint planted cliques, we turn to more complicated problems.
We ask: can our method identify relevant higher-order interactions when the data is (plausibly) more structured?
To answer this question, we use empirical bipartite networks and create higher-order networks, by representing the bipartite networks as hypergraphs $H$~\cite{newman2018networks}.
We then project the hypergraphs with Eq.~\eqref{eq:likelihood}, and attempt to recover the planted higher-order interactions in $\mathcal{G}(H)$ with our method.

\begin{figure}
    \null\hspace{-0.1\linewidth}
    \includegraphics[width=0.9\linewidth]{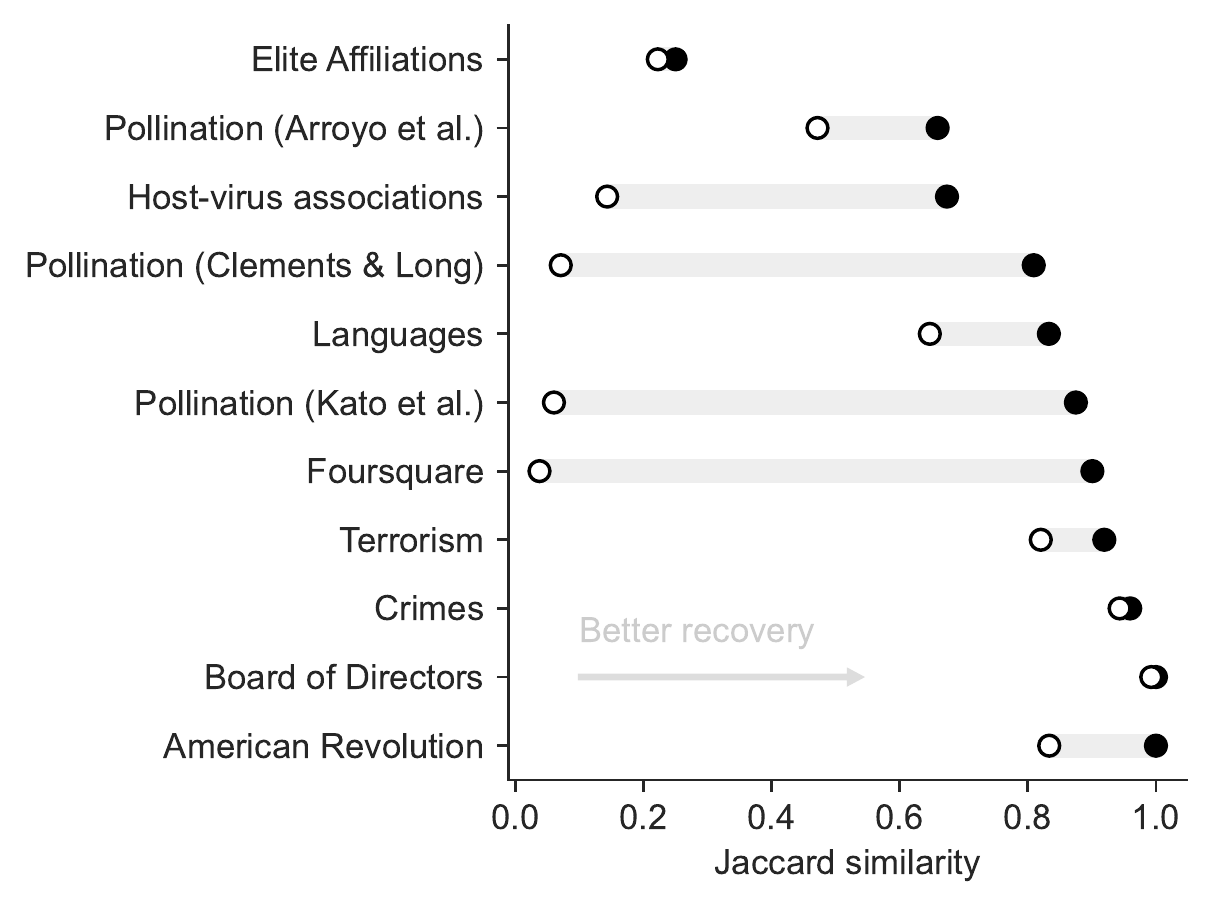}  
    \caption{
    \textbf{Quality of the planted interaction reconstruction in projected bipartite networks.}
    Empty symbols show to the Jaccard similarity of the higher-order interactions reconstructed  with the maximal clique decomposition.
    Filled symbols depict the same quantity for the reconstruction given by our method (best model fit).
    Our method improves on the baseline in every case.
    Detailed numerical results are reported in Supplementary Table 1.
    }
    \label{fig:bipartite_reconstruction}

\end{figure}

In Fig.~\ref{fig:bipartite_reconstruction}, we report the results of this experiment for 11 hypergraph constructed with empirical networks datasets~\cite{barnes2017structural,arroyo1985community,olival2017host,clements1923experimental,kunegis2013konect,kato1990insect,yang2013fine,gerdes2014assessing,university1991st,seierstad2011few,kunegis2013konect}.
(See also Supplementary Table~1 for a detailed numerical account of the results.)
The figure depicts the accuracy of the reconstruction, as quantified a Jaccard similarity $J$ defined as the number of hyperedges found in both the original and the reconstructed hypergraph, divided by the number of hyperedges found in either of them.
A similarity of $J=1$ denotes perfect agreement, while $J=0$ would mean that the hypergraphs share no hyperedges.
(When computing $J$, we ignore duplicate hyperedges since they are impossible to distinguish from the projection. For instance, if the board of many companies comprises the exact same directors, than we encode their association with a single hyperedge.)

To obtain a baseline, we also attempt a reconstruction by identifying the maximal cliques of the projected graph to hyperedges---a maximal clique reconstruction.
We find that the reconstruction given by our method is good but imperfect, which is expected as the problem is under-determined.
However, we also find that our method systematically outperforms the maximal clique decomposition, often by a sizable margin. 
In many cases, the maximal clique decomposition recovers nearly none of the common interactions, whereas our method reconstructs the interactions to a great extent.

\subsection{Detailed case study of higher-order interactions in an empirical network}
\label{subsec:football}

To understand why our method works well in practice,
it is useful to  analyze a small empirical dataset in details. 
For this example, we will consider the well-known football network~\cite{girvan2002community}.
The nodes of this network represent teams playing in Division I-A of the NCAA (now the NCAA Division I Football Bowl Subdivision), and two teams are connected if they played at least one game during the regular season of Fall 2000.
The relationships between teams are viewed through the lens of pairwise interactions, but higher-order phenomena shape the system.
For example, the teams of a conference all play each other during a season.
Other higher-order phenomena such as geography  also intervene: teams in different conferences are likely to meet during the regular season if they are in close-by states.
There might also be more subtle phenomena like historical rivalries that survived conference changes.
Which of these higher-order organizing principles best determine the structure is not that clear, so there is no single natural bipartite representation of the system---it is best to work with the projected network and let the data guide us.

\begin{figure}
\includegraphics[width=\linewidth]{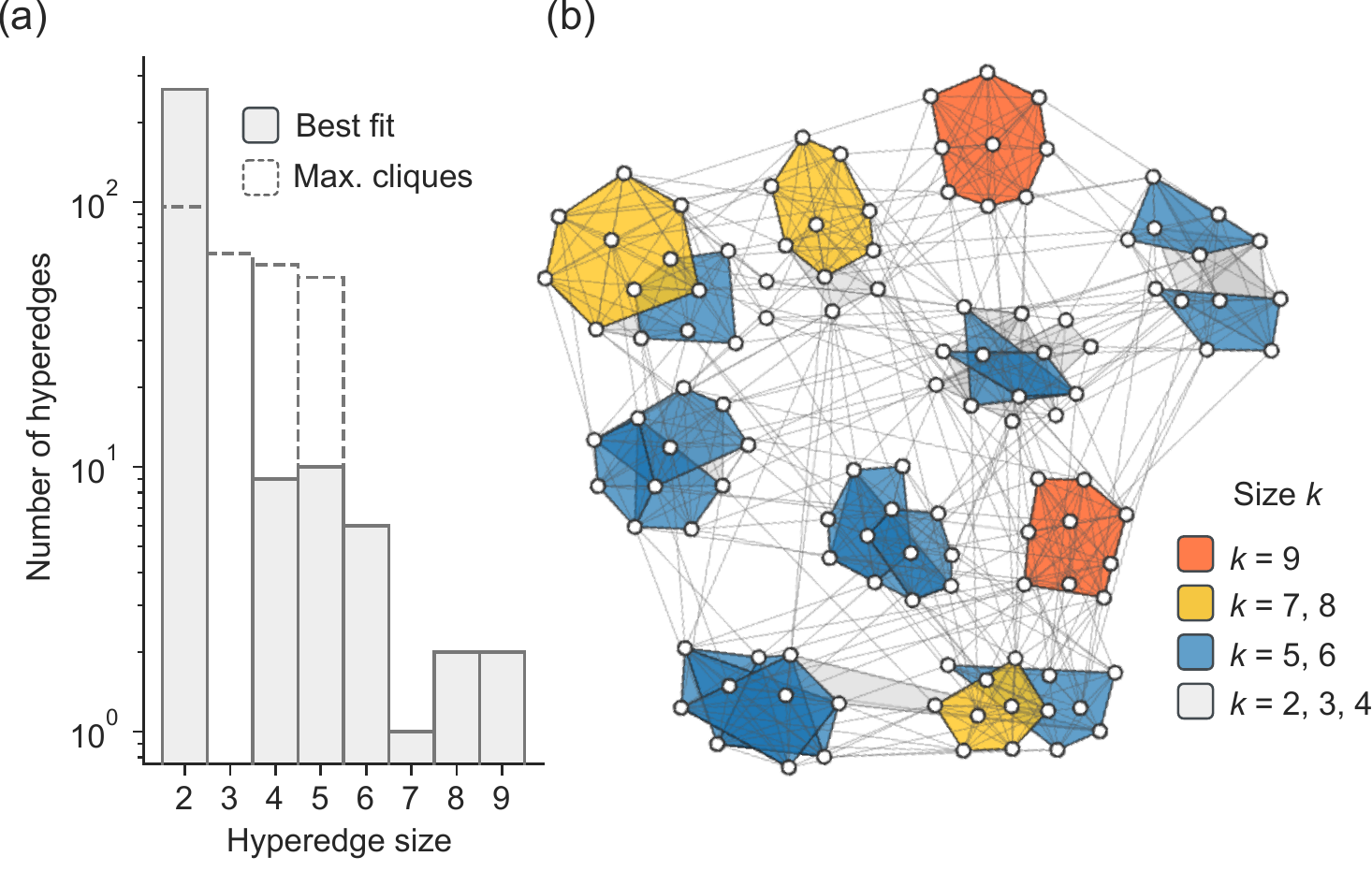}
    \caption{
    \textbf{Higher-order interactions uncovered  in the network of American football games}.
    \textbf{(a)} Size distribution of the hyperedges of the hypergraph $H^*$ that maximizes the posterior probability or, alternatively, minimizes the description length of the network of football games $G$.
    Also shown is the distribution of hyperedge sizes for the hypergraph constructed by assigning a hyperedge to every maximal clique of $G$ (dashed histogram).
    \textbf{(b)} Visualization of the hyperedges present in $H^*$, color-coded by size.
    }
    \label{fig:football_MAP}
\end{figure}

\subsubsection{Best model fit}

In Fig.~\ref{fig:football_MAP} we show the interactions that our method uncovers when we look for the single best higher-order description $H^*$.
We find a large number of interactions that are not pairwise: $30$ of the hyperedges of $H^*$ involve more than $2$ nodes.

The higher-order interactions uncovered by our method are not merely the maximal cliques of $G$ (see Fig.~\ref{fig:football_MAP}a).
We argue that interlocked maximal cliques---cliques that share edges---are the reason why these descriptions differ.
When two maximal cliques interlock, the hypergraph constructed directly from maximal cliques contains two overlapping hyperedges.
This choice is wasteful from a compression perspective: the edges in the intersection of the two cliques are part of two hyperedges, and therefore contribute twice to the description length $\Sigma=-\log P(H)$.
Our method instead looks for a more parsimonious description of the data.
In doing so, it can identify trade-offs and, for example, represent one of the two cliques as a higher-order interaction and break down the other as a series of smaller interactions, thereby avoiding redundancies.
These trade-offs culminate into much better compression: we find a hypergraph $H^*$ with a description length of $2411.3$ bits, which represents a $43.2$\% saving over the description length of the maximal cliques hypergraph ($4246.5$ bits).
The interactions in the optimal hypergraphs do not necessarily map to obvious suspect like sub-divisions or geographical clusters; instead, they interact in non-obvious ways and reveal, for example, that one of the sub-divisions (top left of Fig.~\ref{fig:football_MAP}b) is best described as two interlocking large hyperedges with a few interactions.

\subsubsection{Probabilistic descriptions}

Being Bayesian, our method provides complete estimation procedures, beyond maximum a posteriori estimation.
For example, a quantity of particular interest is the posterior probability that a set of nodes is connected by at least one hyperedge~\cite{young2020bayesian}, once we account for the full distribution over hypergraphs $P(H|G)$.
By computing this probability for all sets of nodes with a non-negligible connection probability, we can encode the probabilistic structure of $H$ in a compact way~\cite{parchas2015uncertain}, with a few probabilities only.

In practice, we evaluate the connection probabilities by generating samples from $P(H|G)$ and counting the samples in which a set of interest is connected by at least one hyperedge (recall that the model defines a distribution over hypergraph with repeated hyperedges).
Mathematically this is computed as  
\begin{equation}
    P(X_{i_1,...i_k}=1|G) = \frac{1}{n} \sum_{\ell=1}^n X_{i_1,...i_k}(H_\ell),
\end{equation}
where $H_1,...,H_n$ are $n$ hypergraph sampled from $P(H|G)$, and where $X_{i_1,..,i_k}(H)= \mathbb{1}_{A_{i_1,..,i_k\geq 1}}$ is a presence/absence variable, equal to 1 if and only if there is at least one hyperedge connecting nodes $i_1,..,i_k$ in hypergraph $H$.

Applying this technique to the Football data, we find that many of the hyperedges of $H^*$ have a presence probability close to 1, even once we account for the full distribution over hypergraphs. 
The hypergraph is not reconstructed with absolute certainty, however.
Observing that probabilities $P(X_{i_1,...i_k}=1|G)$ close to $1$ or $0$ both indicate confidence in the presence/absence of edge, we define a certainty threshold $\alpha$ and classify all hyperedges with existence probabilities in $[\alpha, 1 -\alpha]$ as uncertain.
With a threshold of $\alpha=0.05$, we find $16$ uncertain triangles (hyperedges on 3 nodes), $70$ uncertain edges, and $9$ additional uncertain interactions of higher orders.

\begin{figure}
    \includegraphics[width=\linewidth]{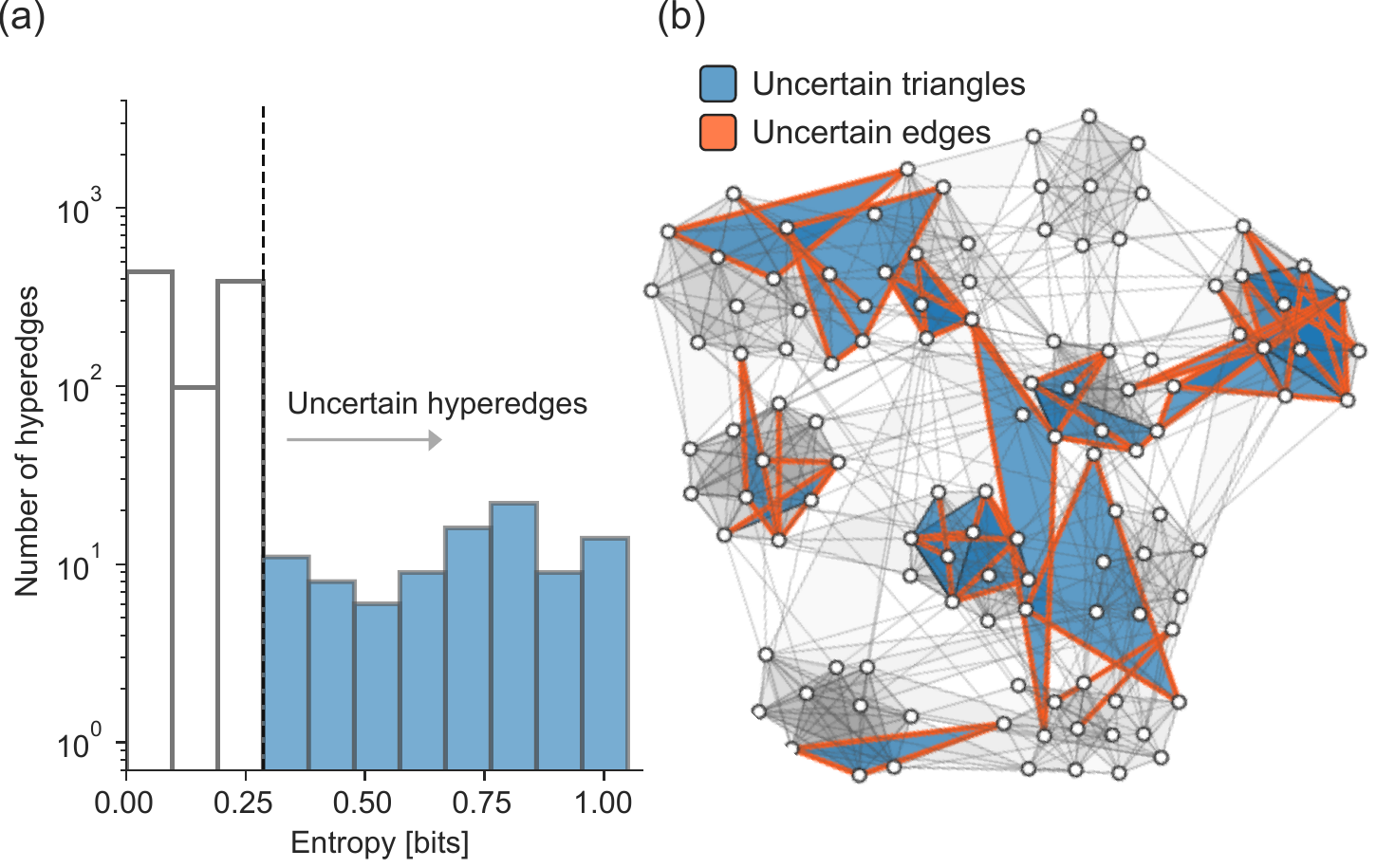}
    \caption{
        \textbf{Uncertain higher-order interactions uncovered in the network of American football games}.
        \textbf{(a)} Distribution of the entropy for the presence/absence of hyperedges.
        The entropy quantifies  the variability of hyperedges: sets of nodes that are connected in nearly all---or almost none---of the samples have low entropy.
        We deem as uncertain a hyperedge that has a probability $p\in [\alpha, 1 - \alpha]$ of being present, with $\alpha=0.05$.
        Note that we only show the entropy for the sets of nodes that were connected at least once in our Monte Carlo samples; a large number of hyperedges, of entropy zero, are never seen in our samples.
        \textbf{(b)} Visualization of the uncertain hyperedges. Uncertain edges are highlighted in orange and uncertain triangles are shown in blue.
        Remaining uncertain interactions are shown in gray.
        All results are computed with $4,000$ Monte Carlo samples from the posterior distribution each separated by $1,000$ complete sweep of the factor graph.
    }
    \label{fig:football_uncertain}
\end{figure}

To go beyond a simple threshold analysis, we compute the entropy of the probabilities $\hat{p}:= P(X_{i_1,...i_k}=1|G)$, defined as
\begin{equation}
    S(\hat{p})=-\hat{p}\log_2 \hat{p} -(1-\hat{p})\log_2(1-\hat{p}).
\end{equation}
The entropy provides a useful transformation of $\hat{p}$ because it grows as $\hat{p}$ moves away from the extremes $\hat{p}=0,1$, with a maximum of $S=1$ at $\hat{p}=1/2$, the point of maximal uncertainty.
The distribution of entropy is shown in Fig.~\ref{fig:football_uncertain}a for the Football data.
The figure shows that while the majority of hyperedges are certain (i.e., their entropy is greater than $S*\approx 0.286$ corresponding to $\alpha=0.05$), making the certainty criterion slightly more stringent would add many more uncertain hyperedges to the ones we already have.

In Fig.~\ref{fig:football_uncertain}b, we show the location of the uncertain hyperedges in $H$.
We observe that these uncertain hyperedges are often concentrated together.
The minimality properties of $P(H|G)$ discussed above can explain these results.
Hypergraphs that have a sizable posterior probability are typically sparse and include as few hyperedges as possible.
But they also need to cover the whole graph, meaning than at every edge of $G$ needs to appear as a subset of at least one hyperedge $H$ (due to the constraint $G=\mathcal{G}(H)$).
Co-located uncertain triangles and hyperedges are hence the result of competing solutions of roughly equal qualities, that cover a specific part of the hypergraph with hyperedges of different sizes.

\subsection{Systematic analysis of higher-order interactions in empirical networks}
\label{subsec:empirical}

\begin{figure*}
    \centering
    \includegraphics[width=0.87\linewidth]{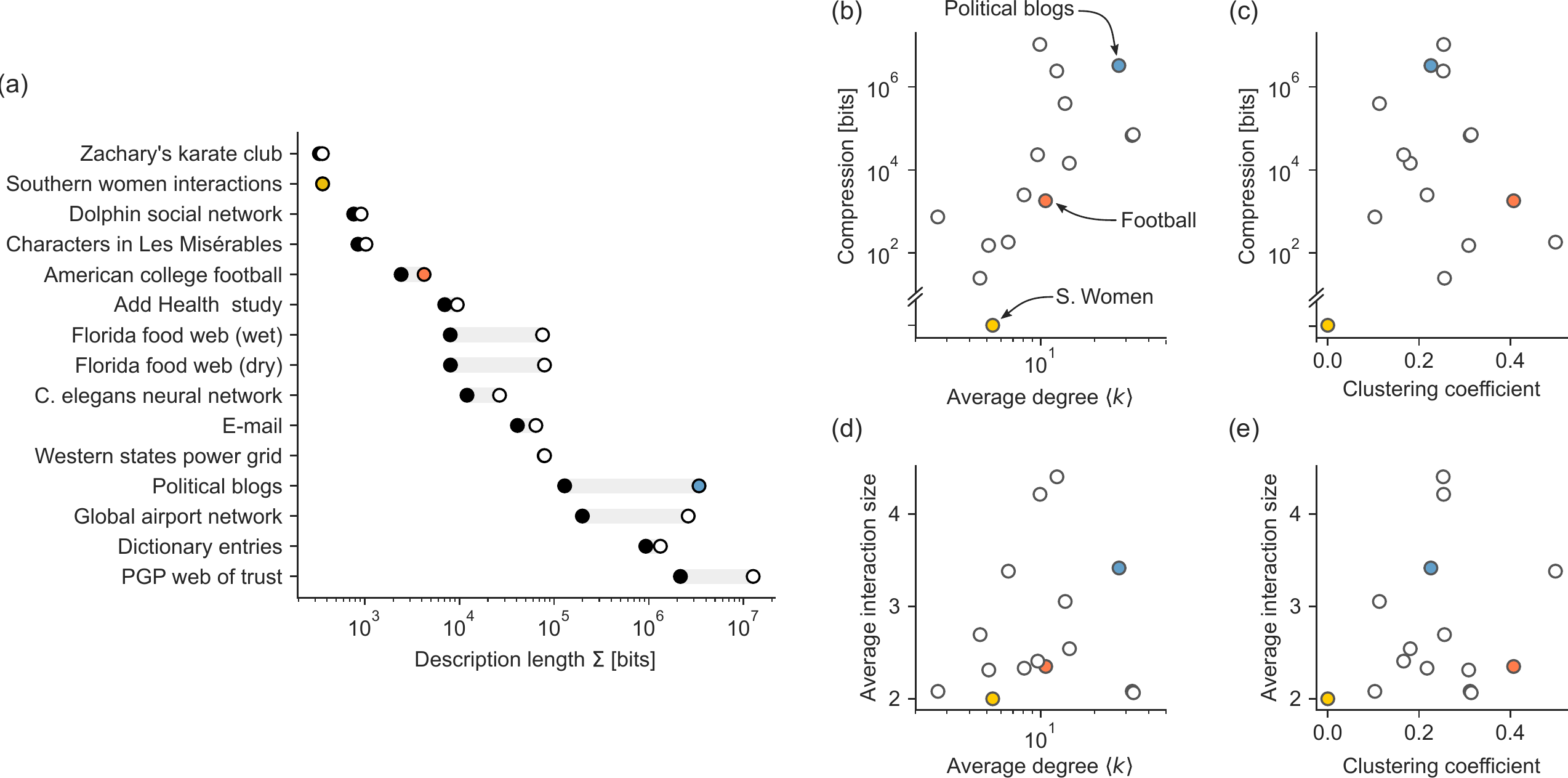}
    \caption{
        \textbf{Higher-order interaction in empirical networks.}
        A few datasets are highlighted with colors: the Southern women interaction data~\cite{davis2009deep} (yellow), the Football data~\cite{girvan2002community} (orange), and the political blogs ~\cite{adamic2005political} (blue).
        \textbf{(a)} Description length of the hypergraph $H$ whose hyperedges are the maximal cliques of the input network $G$, compared with the description length found with our method.
        \textbf{(b,c)} Compression, defined as the difference in description length, as a function of the average degree and clustering coefficient~\cite{newman2018networks}.
        \textbf{(d,e)} Interaction size, averaged over all interactions in $H^*$, as a function of the average degree and clustering coefficient.
        Detailed numerical results are reported in Supplementary Table~2.
    }
    \label{fig:empirical_nets}
\end{figure*}

For our fourth and final example, we apply our method to 15 network data sets, taken from various representative scientific domains and structural classes~\cite{zachary1977information,davis2009deep,lusseau2003bottlenose,knuth1993stanford,girvan2002community,resnick1997protecting,ulanowicz1999network,white1986structure,newman2006modularity,guimera2003self,watts1998collective,adamic2005political,peixoto2014hierarchical,batagelj2002network,richters2011trust}.

For each empirical network in our list, we first search for the hypergraph $H^*$ that maximizes $P(H|G)$, as we have done in our two previous examples.
This search gives us a minimum description length $\Sigma$.
For the sake of comparison, we also compute the description length $\Sigma'$ that we would obtain if we were to use the maximum clique decomposition to construct $H$ naively.
We note that $\Sigma'$ cannot be smaller than $\Sigma$ because it is the description length of the starting point of the MCMC algorithm---at best, the algorithm cannot improve on $\Sigma'$, and we then have $\Sigma=\Sigma'$.
The difference $\Sigma' - \Sigma$ gives the compression factor or, in other words, the number of bits we save by using the best hypergraph instead of a hypergraph of maximal cliques.

In Fig.~\ref{fig:empirical_nets}a, we show the description lengths of  the networks in our collection of datasets.
We observe a broad range of outcomes.
Compression of multiple orders of magnitude is possible in some cases, like with the political blogs data~\cite{adamic2005political} highlighted in blue, while the best description is directly the maximal cliques in others, like with the Southern women interaction data~\cite{davis2009deep} highlighted in yellow.
We find that the average degree correlates with compression (Kendall's $\tau=0.52$); see Fig.~\ref{fig:empirical_nets}b.
This result is expected: the denser a network, the more likely it is that interlocking cliques are present, and therefore that a parsimonious description can be obtained by optimizing over $P(H|D)$.
The average local clustering coefficient $\langle C \rangle$~\cite{newman2018networks} is not correlated with compression, however ($\tau=0.03$); see Fig.~\ref{fig:empirical_nets}c.
Local clustering quantifies the density of closed triangles in the neighborhood of a node and is, as such, a proxy for the density of cliques.
However, as our results show, $\langle C \rangle$ fails to capture the correct type of redundancy necessary for good compression with our model.

We note that clustering nonetheless predicts the absence of compression well:
If $\langle C \rangle=0$, then there are no closed triangles in $G$, and it is impossible to compress the network with our method---there are no cliques, and therefore no higher-order interactions in the data.
The Southern Women~\cite{davis2009deep} falls in this category because it is a bipartite network.

In Figs.~\ref{fig:empirical_nets}d and \ref{fig:empirical_nets}e, we show the size of the higher-order interactions founds by our method, averaged over the hyperedges of $H^*$.
We again observe a wide range of outcomes.
As a sanity check, we can confirm that the incompressible network has an average interaction size of 2.
All hypergraphs are just networks in this case and therefore have no higher-order interactions.
Other datasets yield hypergraphs with large interactions on average, involving as many as 5 nodes the airport network.
The correlation between local properties and interaction size is also weak, though there are some dependencies ($\tau=0.12$ and $\tau=0.09$ for the degree and local clustering, respectively).
These might be partly explained by constrained on the possible values that the average interaction size $\langle s\rangle$ can adopt.
For instance, to have an average size $\langle s \rangle$, a network must have an average degree of at least $\langle s\rangle - 1$.
Likewise, some level of clustering is required to obtain large interactions.

Summarizing these results, we find that some level of compression is always possible, except when the network has no clustering whatsoever.
Furthermore, we find that a high average degree is related to more compression and larger higher-order interactions.
Finally, we find that some minimal level of clustering is necessary for compression, but that results vary otherwise.

\section{Conclusion}
\label{sec:conclusion}

Higher-order interactions shape most relational data sets~\cite{battiston2020networks,torres2020and}, even when they are not explicitly encoded.
In this work, we have shown that it is possible to recover these interactions from data.
We have argued that while the problem is ill-defined, one can introduce regularization in the form of a Bayesian generative model, and obtain a principled recovery method.

The framework we have presented is close in spirit to precursors who have used generative model to find small patterns in networks, so it is worth pointing out where it differs, both in its methodological details and philosophical underpinning.
Closely related work include that of Wegner~\cite{wegner2014subgraph}, who used a notion of probabilistic subgraphs covers to induce distributions over possible decomposition in motifs, and more recent works in graph machine learning that solve graph compression by decomposing the network in small building blocks \cite{koutra2014vog,liu2018reducing}.
Unlike these authors, however, we focused on higher-order interactions, so we considered decompositions in hyperedges rather than in general motif grammars.
We also differ on a methodological ground: we embraced the complexity of the problem and proposed a fully Bayesian method that can account for the multiplicity of descriptions, in contrast with the greedy optimization favored in Ref.~\cite{wegner2014subgraph,koutra2014vog,liu2018reducing}.
As a result of these methodological choices, our work is perhaps closest to that of Williamson and Tec~\cite{williamson2018random}, who also solved a similar problem from using Bayesian nonparametric techniques~\cite{williamson2018random}, and view a network as collections of overlapping cliques.
Unlike these authors, however, we have formalized network data as uncorrupted; in our framework, latent higher-order interactions always show up in network data as fully connected cliques.
In contrast, they think of this process as noisy, so latent higher-order interactions can translate into relatively sparsely connected sets of nodes.
Their proposed methods thus bear a resemblance to community detection techniques that formalize communities as noisily measured cliques~\cite{davis2008clearing,airoldi2008mixed,barber2012clique,xie2013overlapping,verzelen2015community,fortunato2010community}.

The method we have proposed here is undoubtedly one of the simplest instantiations of the broader idea of uncovering higher-order interactions in empirical relational data.
There are many ways in which one could expand on the method.
On the modeling front, for example, it would be worthwhile to study the interplay of the projection component $P(G|H)$ of Eq.~\eqref{eq:likelihood} and inference: can it be defined in a way that does not turn higher-order interaction discovery into overlapping
community detection? 
The hypergraph prior, too, will have to be expanded as the Poisson Random Hypergraphs Model (PRHM) we have used is pretty simple.
Interesting models could include degree heterogeneity as part of the reconstruction~\cite{peixoto2020latent,stasi2014beta,chodrow2020configuration}, or community structure~\cite{chodrow2021hypergraph}.
One could also envision a simplicial analog to these models, leading to probabilistic simplicial complex recovery~\cite{young2017construction,courtney2016generalized}. 
Finally, it would be interesting to explore the connection between different forms of regularizations that make the problem well-defined.

On the technical front, it will be interesting to see whether more refined MCMC methods can lead to more robust convergence and faster mixing.
Our proposed move-set is among the simplest one can propose for the problem and could be improved.
Another interesting avenue of research will be to harness the known properties of $P(H|G)$ to construct efficient inference algorithms, and perhaps connect the method to algorithms in the graph theory of clique covers.

The higher-order interaction data we need to inform the development of higher-order network science~\cite{battiston2020networks} are often inaccessible.
Our methods provide the tools needed to extract higher-order structures from much more accessible and abundant relation data.
With this work, we hope to have shown that moving to principled techniques is possible, and that \textit{ad hoc} reconstruction methods should be avoided, in favor of those based on information-theoretic parsimony and statistical evidence.

\section*{Data availability}
The network data that support the findings of this study are available online in the Netzschleuder network repository~\cite{peixoto_netzschleuder_2020}.

\section*{Code availability}

A fast implementation of the Markov Chain Monte Carlo algorithm described in this study is freely available as part of the \texttt{graph-tool} Python library~\cite{peixoto2014graph}.

\acknowledgments{
This work was funded in part by the James S. McDonnell Foundation (JGY), the Sanpaolo Innovation
Center  (GP), and the Compagnia San Paolo via the ADnD project (GP).
We thank Huan Wang, Yanbang Wang,  Kyuhan Lee, and  Guillaume St-Onge for useful comments.
\\
}

\end{document}